\documentclass[12pt]{iopart}
\usepackage{iopams}
\usepackage{graphicx}
\usepackage{bm}
\usepackage{color}

\newcommand* {\ee}{\ensuremath{\mathrm{e}}}

\graphicspath{{./}{./EPS/}}

\begin{document}

\title{Signatures of tunable Majorana-fermion edge states}

\author{Rakesh P.\ Tiwari$^1$, U. Z\"ulicke$^2$ and C.\ Bruder$^1$}

\address{$^1$ Department of Physics, University of Basel, Klingelbergstrasse 82, CH-4056 Basel, Switzerland}

\address{$^2$ School of Chemical and Physical Sciences and MacDiarmid Institute for Advanced Materials
and Nanotechnology, Victoria University of Wellington, PO Box 600, Wellington 6140, New Zealand}

\ead{rakesh.tiwari@unibas.ch}

\begin{abstract}
  Chiral Majorana-fermion modes are shown to emerge as edge
  excitations in a superconductor--topological-insulator hybrid
  structure that is subject to a magnetic field. The velocity of this
  mode is tunable by changing the magnetic-field magnitude and/or the
  superconductor's chemical potential. We discuss how
  quantum-transport measurements can yield experimental signatures of
  these modes. A normal lead coupled to the Majorana-fermion edge
  state through electron tunneling induces resonant Andreev
  reflections from the lead to the grounded superconductor, resulting
  in a distinctive pattern of differential-conductance peaks.
\end{abstract}

\pacs{73.20.At, 73.25.+i, 74.45.+c, 73.50.Jt}

\maketitle

\section{Introduction}
\label{sec:introduction}

The possibility to create and study Majorana quasiparticles in
condensed-matter systems has become a focus of intense attention in
recent years~\cite{wil09,ali12,lei12,bee13,fra13,stan13}.  Spatially
localized versions of such excitations have been predicted to exist,
e.g., in the $\nu=5/2$ quantum-Hall state~\cite{moo91,ste10}, $p$-wave
superconductors~\cite{iva01} such as strontium ruthenate~\cite{mac03,sar06,tew07},
and semiconductor-superconductor
heterostructures~\cite{sau10,sau10a,ali10}.  Zero-bias conductance
anomalies~\cite{sen01,bol07,law09,fle10} associated with Majorana
quasiparticles have been measured
recently~\cite{mou12,das12,den12,fin13} (see, however, Ref.~\cite{lee14}), and observations of an
unconventional Josephson effect~\cite{kit01} mediated by these
excitations have also been reported~\cite{rok12,wil12}.

Chiral Majorana modes can also be realized as edge states in hybrid
structures formed from a topological insulator~\cite{has10,qi11} (TI),
an $s$-wave superconductor (S), and a ferromagnetic
insulator~\cite{fu08,fu09,tan09}. The requirement of broken
time-reversal symmetry and gapped excitation spectrum for the surface
states in the TI is fulfilled by proximity to a ferromagnetic
insulator~\cite{fu09,tan09} or Zeeman splitting due to a magnetic
field~\cite{fu08}. In an alternative realization, Landau quantization 
of the surface
states' orbital motion in a uniform perpendicular magnetic field could
be the origin of the gap and breaking of time-reversal
symmetry~\cite{tiw13}. This setup avoids
materials-science challenges associated with the fabrication of the
hybrid structures involving three different kinds of materials and has
new features enabling the manipulation of the Majorana
excitation's properties. It was shown in Ref.~\cite{tiw13} that the
velocity of the chiral Majorana mode (CMM) can be tuned by changing
the magnitude of the external magnetic field. In this article, we
further explore the properties of this tunable Majorana excitation and
its signatures in typical transport experiments. We show that a normal
lead coupled to this tunable CMM through electron tunneling would
measure a differential conductance that oscillates as the magnitude of
the external magnetic field is changed. The oscillations in the
conductance arise due to the velocity tunability of this
CMM. Recently, it has been proposed that this velocity tunability
could be used for adiabatic quantum pumping induced by Majorana
fermions revealing the chiral nature of these modes~\cite{alo13}.
Some crucial aspects of interferometry with these chiral Majorana modes are highlighted in Ref.~\cite{par13}.

The remainder of this paper is organized as follows. In
Sec.~\ref{sec:snh}, we describe the specific interferometer-like
sample geometry where Landau-quantization-induced CMMs could be probed
by quantum-transport experiments. In Sec.~\ref{sec:mod}, we discuss
the electronic properties of a S--TI interface that is subject to a
magnetic field, showing the existence of Andreev edge
states~\cite{akh07,hop00} and emergence of a Majorana excitation
amongst them~\cite{tiw13}. In Sec.~\ref{sec:eme}, we apply the results
of Sec.~\ref{sec:mod} to elucidate the properties of the CMM that is
present in the particular sample geometry considered here. We then
present numerical results for the conductance of the system in
Sec.~\ref{sec:con}, revealing the signatures of the CMM.  The final
Section~\ref{sec:sum} gives a brief discussion of experimental
parameters and the conclusions.

\begin{figure}
\begin{center}
\includegraphics[scale=0.3]{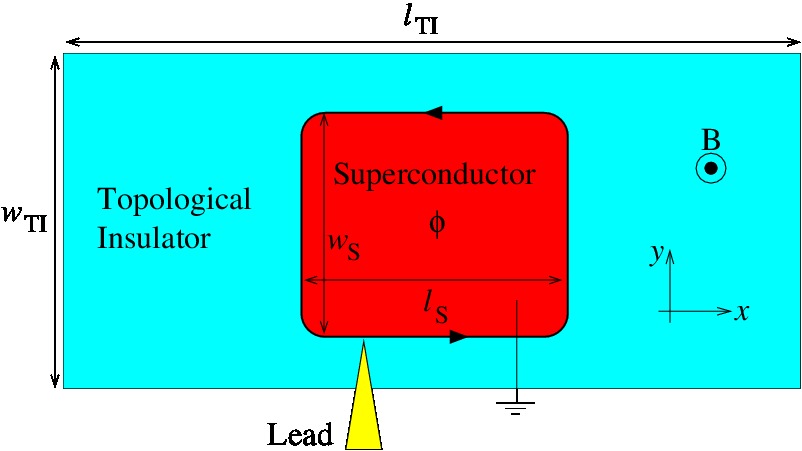}
\caption{Schematics of the sample layout. The red region on the
  surface of a topological insulator denotes the S region that is
  induced via proximity effect due to a planar contact with an
  $s$-wave superconductor. The chiral Majorana mode (CMM) emerges at
  the boundary of this S region. The yellow triangle denotes an ideal
  normal electronic lead that is coming from the $z$--direction and couples
  to the CMM via electron tunneling. The superconductor is grounded,
  and a perpendicular external magnetic field of magnitude $B$ is
  applied along the $z$--axis.}
\label{fig:setup}
\end{center}
\end{figure}

\section{Setup of the superconductor--topological-insulator 
hybrid structure}
\label{sec:snh}

Figure~\ref{fig:setup} illustrates the proposed sample geometry. The
massless-Dirac-like surface states of a bulk three-dimensional TI
material occupy the $xy$ plane. A planar contact with a superconductor
(indicated by the red region in the center of the TI surface) induces
a pair potential for this part of the TI surface (henceforth called
the S region), while a uniform perpendicular magnetic field ${\bf B} =
B\hat{{\bf z}}$ is present in the rest of the surface (the N region of
the TI surface). We assume the magnetic field to create a finite
number of vortices in the S part and neglect Zeeman splitting
throughout. (A more detailed theoretical treatment of
effects due to the finite magnetic field in S
can be developed along the lines of previous
studies~\cite{zue01,gia05}.) The dimensions of the S and N regions are
denoted by $l_S \times w_S$ and $l_{TI} \times w_{TI}$
respectively. We assume
\begin{equation}
\{l_{TI}, w_{TI}, l_S, w_S, l_{TI}-l_S, w_{TI}-w_S\} \gg l_B \:, 
\label{eq:lengthscales}
\end{equation}
where $l_B=\sqrt{\hbar/|e B|}$ denotes the magnetic length.  The
cyclotron motion of charge carriers in the N region hybridizes with
Andreev reflection from the interface with the S region.  As shown in
Ref.~\cite{tiw13}, this results in the formation of chiral
Dirac-Andreev edge states, similar to the ones discussed previously
for an S--graphene hybrid structure~\cite{akh07}. The quantum
description of these edge channels reveals that one of them is
associated with a chiral Majorana fermion mode with tunable velocity
and guiding-center-dependent electric charge.

We now consider the situation where this CMM is tunnel-coupled to an
ideal normal electronic lead coming from the $z$--direction, e.g., as shown
in Fig.~\ref{fig:setup}. The interior of the S region may accommodate
$n_v$ number of vortices, each permitting a magnetic flux quantum
threading through and containing a Majorana bound state. The chiral
Majorana fermion traveling around the boundary of the S region picks
up a phase that contains information about the number of vortices and
the velocity of the mode before scattering into the normal
lead. Before calculating the differential conductance in our setup, we
compute its excitation spectrum and demonstrate the presence of the
CMM along the boundary of the S region.

\section{Theoretical description of the S--TI interface in a 
magnetic field}
\label{sec:mod}

Single-particle excitations in the S--N heterostructure made from TI
surface states can be described by the Dirac-Bogoliubov-de~Gennes
(DBdG) equation~\cite{akh07,fu08,tiw13}
\begin{equation}
\left( \begin{array}{cc}
H_{D}({\bf r})-\mu & \Delta({\bf r})\,\sigma_0 \\ 
\Delta^{\ast}({\bf r})\, \sigma_0 &
\mu-\mathcal{T} H_{D}({\bf r})\, \mathcal{T}^{-1} \end{array}\right) 
\Psi({\bf r}) = \varepsilon \Psi({\bf r}) \: ,
\label{eq:hamiltonian} 
\end{equation}
where the pair potential $\Delta({\bf r})=\ee^{i\theta}\Delta_0$ is
finite only in the S region, $\sigma_0\equiv\mathbb{I}_{2\times 2}$
denoted the two-dimensional identity matrix, and $H_{D}({\bf r}) =
v_{F} [{\bf p} + e{\bf A}({\bf r})] \cdot {\bm \sigma}$ is the
massless-Dirac Hamiltonian for the TI surface states. The position
${\bf r}\equiv (x, y)$ and momentum ${\bf p}\equiv -i\hbar
(\partial_x, \partial_y)$ are restricted to the TI surface. ${\bm
  \sigma}$ is the vector of Pauli matrices acting in spin
space. Furthermore, $\mathcal{T}$ denotes the time-reversal operator,
$-e$ the electron charge, and ${\bf A}$ the vector potential
associated with the magnetic field ${\bf B}={\bf \nabla}\times{\bf
  A}$. The excitation energy $\varepsilon$ is measured relative to the
chemical potential $\mu$ of the superconductor, with the absolute zero
of the energy set to be at the Dirac (i.e., neutrality) point of the
TI surface states. The wave function $\Psi$ in 
Eq.~(\ref{eq:hamiltonian}) is a spinor in Dirac-Nambu space, which can
be expressed explicitly in terms of spin-resolved amplitudes as
$\Psi=(u_{\uparrow},
u_{\downarrow},v_{\downarrow},-v_{\uparrow})^{T}$. As we will confirm 
later, the zero-energy solution of the DBdG equation
localized at the boundary between the S and the N regions constitutes
the chiral Majorana excitation in our system.

To describe the uniform perpendicular magnetic field in the N region,
we adopt the Landau gauge ${\bf A}=B\, x\, \hat{{\bf y}}$. To
explicitly verify the presence of these Majorana excitations at the
boundary of our S-N heterostructure, we restrict ourselves to the
right boundary of the system. Assuming that the various lengths in our
system satisfy Eq.~(\ref{eq:lengthscales}), we model this right
boundary as a one-dimensional edge ($x=0$) between two half
planes. The left half-plane ($x<0$) represents the S region and the
right half plane ($x>0$) represents the N region. Then the momentum
$\hbar q$ parallel to the interface (i.e., in $\hat{{\bf y}}$
direction) is a good quantum number of the DBdG Hamiltonian, and a
general eigenspinor is of the form
\begin{equation}
\Psi_{n q}({\bf r}) = \ee^{ \frac{i\theta}{2}\, 
\sigma_0\otimes\tau_z} \,\, \ee^{i q y} \, \Phi_{nq}(x) \: .
\end{equation}
Here $\tau_z$ is a Pauli matrix acting in Nambu space, $\sigma_0$ the
identity in spin space, and $n$ enumerates the energy (Landau) levels
for a fixed $q$. The spinors $\Phi_{nq}(x)$ are solutions of the
one-dimensional (1D) DBdG equation ${\mathcal H}(q) \, \Phi_{n q}(x) =
\varepsilon\, \Phi_{nq}(x)$, with
\begin{eqnarray}
{\mathcal H}(q) &=& \hbar v_{\rm{F}} \left\{ \sigma_x\otimes 
\tau_z (-i) \partial_x + \sigma_y
\otimes \left[ \tau_z \, q + \tau_0\, \frac{e B}{\hbar}\, x \,
  \Theta(x) \right]\right\} \nonumber \\ && \hspace{1cm} 
-\mu \, \sigma_0\otimes\tau_z
 + \Delta_0\, \Theta(-x) \,\sigma_0\otimes\tau_x \:,
\end{eqnarray}
where the $\tau_j$ are Pauli matrices acting in Nambu space, and
$\tau_0$ is the identity matrix in Nambu space. The spectrum of
Landau-level (LL) eigenenergies $\varepsilon_{n q}$ and the explicit
expressions for $\Phi_{nq}(x)$ in the N and S regions can then be obtained 
by demanding the continuity of the wavefunction at the S--N interface. 

\begin{figure}
\begin{center}
\includegraphics[width=0.7\columnwidth,angle=270]{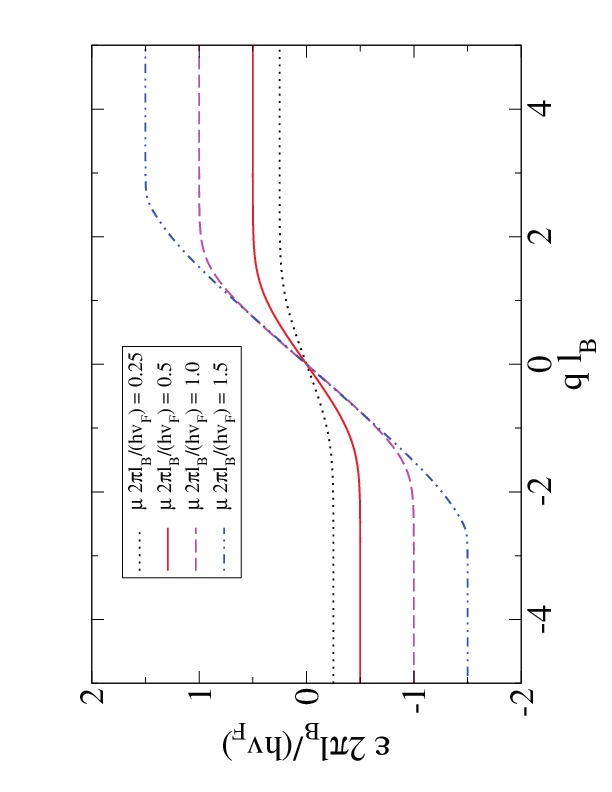}
\caption{Dispersion relation
  $\varepsilon_{nq}$ of single-particle excitations at the S--N
  junction on a TI surface that is subject to a strong perpendicular
  magnetic field. Only the zeroth Landau level is shown for various 
  values of $\mu$ as indicated in the legend. The magnitude of the
  pair potential is $\Delta_0 = 5 \hbar v_{{F}}/l_B$, and 
  $l_B\equiv\sqrt{\hbar/|eB|}$ denotes the magnetic length.}
\label{fig:envsq}
\end{center}
\end{figure}
The functional form of the solutions to this 1D DBdG equation decaying
for $x\rightarrow \infty$ (in the region $x>0$) can be expressed as 
\begin{equation}
\Psi(x,y)=e^{iqy}\left(\begin{array}{c}
                             -i C_e e^{-(x + q)^2/2}(\mu +\varepsilon)H_{(\mu +\varepsilon)^2/2 -1}(x +q) e^{i\phi/2} \\
                             C_e e^{-(x + q)^2/2} H_{(\mu +\varepsilon)^2/2}(x +q) e^{i\phi/2} \\
                             C_h e^{-(x - q)^2/2} H_{(\mu -\varepsilon)^2/2}(x -q) e^{-i\phi/2} \\
                             -i C_h e^{-(x - q)^2/2}(\mu -\varepsilon)H_{(\mu -\varepsilon)^2/2 -1}(x -q) e^{-i\phi/2}
                            \end{array}
\right),
\label{eq:wfn}
\end{equation}
with $H_{\alpha}(x)$ denoting the Hermite function~\cite{akh07}.  The
complete solution is then obtained from the requirement of
particle-current conservation across the interface.  Similar to
Ref.~\cite{akh07} we find
\begin{equation}
C_e=\frac{-i\Delta_0C_h(\mu-\varepsilon)
H_{(\mu -\varepsilon)^2/2 -1}(-q)}{H_{(\mu +\varepsilon)^2/2}(q)
\varepsilon+(\mu +\varepsilon)H_{(\mu +\varepsilon)^2/2 -1}(q) 
\sqrt{\Delta_0^2-\varepsilon^2}}\:, 
\label{eq:cech}
\end{equation}
and the dispersion relation is given by the solutions of
\begin{eqnarray}\label{eq:eigenvalue}
&& f_{\mu+\varepsilon}(q)-f_{\mu-\varepsilon}(-q)=\frac{\varepsilon \left[ f_{\mu+\varepsilon}(q)
f_{\mu-\varepsilon}(-q)+1\right]}{\sqrt{\Delta_0^2-\varepsilon^2}} \:, \\
&& \textrm{where} \,\, f_\alpha(q) = \frac{H_{\alpha^2/2}(q)}{\alpha  H_{\alpha^2/2 -1}(q)} \:.
\end{eqnarray}

The solutions $\varepsilon_n(q)$ of Eq.~(\ref{eq:eigenvalue}) can be
labeled with a (LL) index $n=0, \pm 1, \pm 2, \cdots$. Figure
~\ref{fig:envsq} shows the zeroth LL ($n=0$) for $\mu=0.25\,\hbar
v_F/l_B$, $0.5\,\hbar v_F/l_B$, $1.0\,\hbar v_F/l_B$, and $1.5\,\hbar
v_F/l_B$ for $\Delta_0=5 \hbar v_F/l_B$. For $\mu=0$, we obtain the
familiar dispersionless LLs at 0, $\pm \sqrt{2}\,\hbar v_F/l_B$,
$\cdots$~\cite{akh07}. Away from the edge the various LLs saturate at
$\sqrt{2}(\hbar v_F/l_B)$sgn$(n)\sqrt{\mid n \mid} - \mu$ for $q
\rightarrow-\infty$ and $\sqrt{2}(\hbar v_F/l_B)$sgn$(n)\sqrt{\mid n
  \mid} + \mu$ for $q \rightarrow\infty$. This suggests that an
interesting regime can be reached by increasing $\mu$, so that
$n=\pm1$ levels start contributing to the low-energy excitations of
the system (as shown in Ref.~\cite{tiw13}). When the chemical
potential $\mu$ is finite (as measured from the charge-neutrality
point of the free Dirac system), the LLs acquire a dispersion around
$q=0$ that signals the existence of Andreev edge
excitations~\cite{hop00,akh07}.

For the special case of $\varepsilon=0$ and $q=0$, we obtain
$C_h=iC_e$, and this zero-energy state can then be expressed as
\begin{equation}
\Psi_{00}(x)=C_e e^{-\frac{x^2}{2}}\left(\begin{array}{c}
            -i \mu H_{\mu^2/2 -1}(x)e^{i\phi/2} \\
              H_{\mu^2/2}(x)e^{i\phi/2} \\
            i  H_{\mu^2/2}(x)e^{-i\phi/2} \\
              \mu H_{\mu^2/2 -1}(x)e^{-i\phi/2}
           \end{array}\right).
\label{eq:majn_b}
\end{equation}

The particle-hole-conjugation operator is given by $\Xi=\sigma_y\tau_y
\mathcal{K}$ ~\cite{fu08}, where $\sigma_j$ and $\tau_j$ are again the Pauli
matrices acting on spin and particle-hole space respectively, and
$\mathcal{K}$ symbolizes complex conjugation. Straightforward verification
establishes $\Xi\Psi_{00} = -i\Psi_{00}$, and $\Xi\Xi\Psi_{00}=\Psi_{00}$.
Hence the state $\Psi_{00}$ is a Majorana fermion.

\section{Chiral Majorana mode in our sample geometry}
\label{sec:eme}

\begin{figure}
\begin{center}
\includegraphics[scale=0.3]{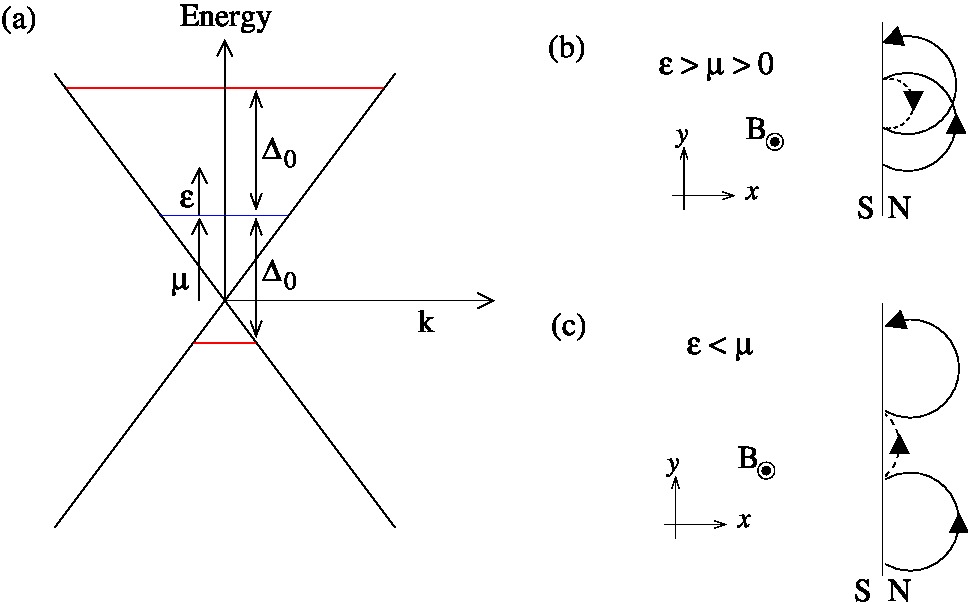}
\caption{Semiclassical picture of Dirac-Andreev edge
  states~\cite{akh07}. (a) Conical dispersion relation representing
  the surface states of the TI, consisting of a top cone (conduction
  band) and a symmetric bottom cone (valence band), which touch at the
  Dirac point. A proximity-induced superconducting gap of $2\Delta_0$
  opens up at the chemical potential $\mu$.  (b) If the excitation
  energy $\varepsilon > \mu$, the electron and the Andreev-reflected
  hole belong to the conduction and valence band, respectively, and
  the edge channel is not chiral.  (Solid and dashed lines represent
  electron and hole trajectories, respectively.)  (c) If $\varepsilon
  < \mu$, electrons and Andreev-reflected holes belong to the same
  band, and their skipping orbits form a chiral Andreev edge channel.
}
\label{fig:cartoon}
\end{center}
\end{figure}

A general symmetry property~\cite{deg89} of the DBdG equation mandates
that, for any eigenstate $\Psi_{nq}$ with excitation energy
$\varepsilon_{nq}$, its particle-hole conjugate $\Xi\,\Psi_{nq}$ in
Nambu space is also an eigenstate and has excitation energy
$-\varepsilon_{nq}$.  This symmetry implies that the zero-energy state
with quantum numbers $n=0$ and $q=0$ is its own particle-hole
conjugate in our system, thus exhibiting the defining property of a
Majorana fermion~\cite{wil09,ali12}. While the Majorana state
$\Psi_{00}({\bf r})$ has a localized spatial profile in the direction
perpendicular to the S--N junction (i.e., along the $x$--axis), it is
completely delocalized in the direction parallel to the S--N
interface. Similarly, one can show the existence of the chiral
Majorana edge excitation all around the boundary of the S region, as
is required by the particle-hole symmetry at zero energy. Thus the CMM
encloses the entire S region that is created on the TI surface via
the proximity effect. In principle, the S region may include $n_v$
vortices, each supporting a Majorana bound state (MBS) at zero
energy. The coupling between these MBSs and the CMM discussed above
decays exponentially as a function of their spatial separation. Thus
the presence of vortices will not affect the Majorana edge mode in the
typical situation where the vortices are located far from the edge.

The chiral Majorana edge excitation is characterized by its velocity
$v_M$. As the chemical potential approaches zero, the edge dispersion
of the zeroth LL flattens out, implying a very small Majorana-mode
velocity ($v_M\sim0$). On the other hand, increasing the chemical
potential sharpens that edge dispersion and increases $v_M$ (see
Fig.~\ref{fig:envsq}). In previously considered situations where CMMs
emerge~\cite{fu09}, the Majorana-mode velocity could be adjusted by
changing the magnitude of the magnetization in a ferromagnetic
insulator. However this is rather difficult to perform
experimentally.  Our setup offers a more controllable route to tune
this velocity by changing the magnitude of the external
magnetic field.

The semiclassical cyclotron trajectories of electrons and Andreev
reflected holes for this system are shown in Fig.~\ref{fig:cartoon},
where we consider a finite positive value of
$\mu$. Figure~\ref{fig:cartoon}(a) shows the conical dispersion
relation describing the topologically protected surface states of the
TI. Due to the proximity effect a gap of magnitude $2\Delta_0$ opens up at
$\mu$. First we consider $\Delta_0 > \mu$. In this regime, there exist
two kinds of cyclotron trajectories. In the first case $\epsilon >
\mu$ shown in Fig.~\ref{fig:cartoon}(b), the electron and its
conjugate Andreev-reflected hole are from different (conduction and
valence) bands defined with respect to the Dirac point. The
semiclassical cyclotron trajectories for the electrons and the 
Andreev-reflected holes in this case travel in opposite direction. In the
second case $\epsilon < \mu$, both the electron and its conjugate
Andreev-reflected hole belong to the same (e.g., the conduction) band
and move in the same direction. The chiral modes we discuss belong to
the latter situation and are present even if $\Delta_0 <
\mu$. However, by increasing $\Delta_0$, we can realize a regime with
multiple LLs within the gap. In this regime the coupling between the
nearest LLs enriches the dynamics of the system. The low-energy
excitations (close to the chemical potential) can then be described by
an effective envelope-function Hamiltonian including $n=0$ and $\pm1$,
as discussed in Ref.~\cite{tiw13}.

\section{Conductance of the CMM interferometer}
\label{sec:con}

\begin{figure}
\begin{center}
\includegraphics[angle=0,scale=0.7]{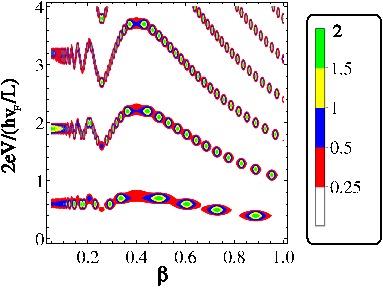} 
\caption{Contour plot of the differential conductance
  $(dI/dV)/(e^2/h)$ as a function of the dimensionless 
  magnetic field denoted by $\beta=\sqrt{\hbar|e B|}v_F/\Delta_0$, and
  $eV$ measured in units of $\pi\hbar v_F/L$. The reflection amplitude
  $r_0$ of the locally coupled normal lead is set to $0.9$, the
  chemical potential is chosen to be $\mu=0.5
  \Delta_0$, and the S region encloses an even number of vortices.}
\label{fig:didvvsphi}
\end{center}
\end{figure}

\begin{figure}
\begin{center}
\includegraphics[angle=0,scale=0.7]{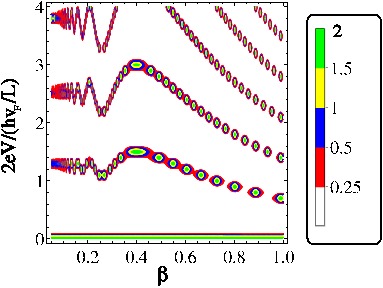} 
\caption{Same as Fig.~\ref{fig:didvvsphi}, but for odd number of vortices.}
\label{fig:didvvsphi2}
\end{center}
\end{figure}

The differential conductance of our setup (as shown in
Fig.~\ref{fig:setup}) can be calculated within the scattering matrix
formalism~\cite{naz09,li12}. Ignoring the coupling between the MBSs
and the chiral Majorana edge state enclosing the S region, the effect
of vortices is to account for an additional contribution $\phi_v=n_v \pi$ to
the total phase $\phi$ that the chiral Majorana edge fermion picks up
after traversing the entire boundary of the S region. Including an
additional phase of $\pi$ coming from the Berry phase and the dynamic
phase, the total phase acquired during a round trip is
\begin{equation}
 \phi=n_v\pi + \pi + \frac{E L}{\hbar v_M} \:,
\end{equation}
where $L$ is the circumference of the S region and $E$ is the energy
of the propagating CMM state.  The normal lead in this setup plays
the role of both the electron and hole lead. The tunneling current
from the lead to the grounded superconductor can be calculated
straightforwardly by working in the Majorana basis \cite{li12}. An
incoming electron or hole from the normal lead can be expressed in
terms of \textit{artificial} Majorana modes $\eta_1$ and $\eta_2$. In
the Majorana basis, only one of the Majorana modes in the normal lead
($\eta_1$ or $\eta_2$) is coupled to the CMM going around the S
region, the other is reflected with amplitude
1~\cite{li12}. Accounting for multiple reflections of the Majorana
mode in the lead, the scattering matrix in the Majorana basis
is~\cite{li12}
\begin{equation}
\mathcal{S}_M=\left(\begin{array}{cc}
                     r_1 & 0 \\
                     0 & r_2
                    \end{array}
\right) \:,
\label{eq:scatmatmaj}
\end{equation}
where $r_2=1$, and
\begin{equation}
 r_1=\frac{r_0 - e^{i\phi}}{1-r_0 e^{i\phi}}\:,
\label{eq:r1}
\end{equation}
with $r_0$ being the local reflection amplitude of $\eta_1$ at the
junction ~\cite{li12}. At zero temperature, the time-averaged current
from the normal lead to the grounded superconductor is~\cite{li12}
\begin{equation}
 I=\frac{e}{h}\int_{0}^{eV}\left(1 - Re[r_1r_2^{\ast}]\right)dE \:.
\label{eq:current}
\end{equation}

We calculate the differential conductance of our setup by numerically
evaluating the velocity of the CMM in the zeroth LL,
\begin{equation}
v_M=\hbar^{-1}\frac{d\varepsilon_n(q)}{dq}\bigg|_{q\rightarrow0} \:. 
\end{equation}
We assume that $v_M$ is constant all around the S region and determine
it from the solution of
Eq.~(\ref{eq:eigenvalue}). Figure~{\ref{fig:didvvsphi}} shows a
contour plot of the differential conductance ($dI/dV$) in units of
$e^2/h$ as a function of the dimensionless magnetic field $\beta=
\sqrt{\hbar|e B|}v_F/\Delta_0$, and $eV$ measured in units of
$\pi\hbar v_F/L$. The chemical potential is chosen to be $\mu=0.5
\Delta_0$, $r_0=0.9$, and we consider an even number of vortices. The
apparent highly nonlinear behavior arises from the Majorana-fermion
induced resonant Andreev reflection (MIRAR)~\cite{law09} in
conjunction with the variation of $v_M$ as the magnitude of magnetic
field changes. The differential conductance is periodic in $eV$, and
the period depends on the CMM velocity. The nonlinear behavior due to
MIRAR along the $eV$ axis was investigated in detail in
Ref.~\cite{law09}. In our setup, $dI/dV$ shows a nonlinear oscillating
behavior as a function of the magnetic-field-dependent parameter
$\beta$ also. The origin of these oscillations can be traced back to
the magnetic-field dependence of $v_M$ as shown in
Ref.~\cite{tiw13}. Recent studies~\cite{alo13} have indicated that
such a velocity modulation could be used as a pump parameter in an
adiabatic-quantum-pumping setup where the pumped current is induced by
the CMM.

We have also calculated the differential conductance for the case of
an odd number of vortices present in the S region, see
Fig.~\ref{fig:didvvsphi2}. In the zero-bias limit,
$dI/dV|_{eV\rightarrow0}\rightarrow 2e^2/h$.  In contrast, for the
case of an even number of vortices shown in Fig.~\ref{fig:didvvsphi},
$dI/dV|_{eV\rightarrow0}\rightarrow 0$. It has been suggested that
this jump in conductance with the parity of the number of vortices is
a clear signature of the Majorana mode~\cite{law09}.

\begin{figure}
\begin{center}
\includegraphics[angle=0,scale=0.7]{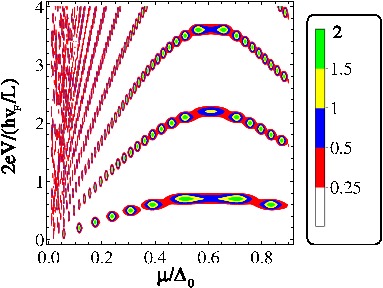} 
\caption{Contour plot of the differential conductance
  $(dI/dV)/(e^2/h)$ as a function of $\mu/\Delta_0$ and $eV$ measured
  in units of $\pi\hbar v_F/L$. The dimensionless measure
  $\beta=\sqrt{\hbar|e B|}v_F/ \Delta_0$ of the magnetic field is set
  to $0.5$. Moreover, the S region encloses an even number of vortices
  and $r_0=0.9$.}
\label{fig:didvvsphi3}
\end{center}
\end{figure}

\begin{figure}
\begin{center}
\includegraphics[angle=0,scale=0.7]{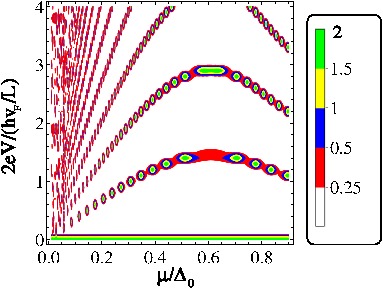} 
\caption{Same as Fig.~\ref{fig:didvvsphi3}, but for odd number of vortices.}
\label{fig:didvvsphi4}
\end{center}
\end{figure}

Along similar lines we calculate the differential conductance for our
setup for a fixed value of magnetic field $\beta =\sqrt{\hbar|e
  B|}v_F/\Delta_0=0.5$, as a function of $\mu$ and $eV$. The results
are shown in Fig.~\ref{fig:didvvsphi3} for an even number of vortices,
and in Fig.~\ref{fig:didvvsphi4} for an odd number of vortices.  In
the zero bias limit, we once again observe a jump in the differential
conductance from $2e^2/h$ to $0$ when the number of vortices changes
from odd to even. We also observe oscillations as the chemical
potential changes which can be traced back to oscillations in
$v_M$~\cite{tiw13}.

\section{Discussion and summary}
\label{sec:sum}

Finally, we would like to comment upon the experimental feasibility of
realizing the device setup suggested here.  Using typical values such
as $\Delta_0=0.7$~meV~\cite{zha11}, $v_F=5 \times
10^5$~m/s~\cite{zha11}, and a magnetic field of $1$~T, we obtain
$\hbar v_F/l_B\sim 10$ meV, implying that the observation of the CMM
discussed here is within the reach of current experimental
efforts. However, realizing a regime with multiple LLs within the gap
necessitates an increase in the proximity-induced superconducting gap
$\Delta_0$. To ensure that a finite temperature $T$ does not
completely smear out the effects we discussed, we have to ensure that
$k_B T \ll \{\Delta_0, \hbar v_F/l_B, \mu \}$, where $k_B$ is the
Boltzmann constant. Therefore, measurements at sub-Kelvin temperatures
will be required for these experiments.

We have ignored disorder in our calculation. Disorder in the system
will result in fluctuations of the Dirac-point energy which will
translate into a spatially varying chemical potential. Majorana
excitations will still exist but be subject to a fluctuating
velocity. Due to the topological nature of the chiral Majorana
edge excitation we expect its largely unimpeded propagation even
under such circumstances. In the limit of strong disorder the
fluctuations in the Majorana mode velocity may suppress
the oscillations described in Figs. \ref{fig:didvvsphi} --
\ref{fig:didvvsphi4}.

In addition to creating the S region via the proximity effect, the planar
contact with a superconducting material on top of the TI surface is likely
to also induce band bending in the TI surface state beneath it. In the likely
case where the resulting potential gradient is smooth on the scale of the
magnetic length, the electronic structure of the Majorana mode discussed
here will be largely unaffected. Furthermore, as normal reflection is
suppressed for states with guiding centers close to the interface because
of Klein tunneling~\cite{cul12}, the Majorana-mode velocity will, in general,
be most dominantly determined by Andreev reflection.

Our theory applies to a situation without vortices or a small number
of them, i.e., for magnetic fields below or just above the first
critical field $B_{\mathrm{c}1}$ of the superconductor. For
$B<B_{\mathrm{c}1}$, there are no vortices in the superconducting
region, and the conductance for this case corresponds to that shown in
Figs.~\ref{fig:didvvsphi} and \ref{fig:didvvsphi3}. For fields just
above $B_{\mathrm{c}1}$, the vortices will be separated by distances
larger than the magnetic penetration depth $\lambda$. For large
$\kappa=\lambda/\xi$, where $\xi$ is the coherence length of the
superconductor, the coupling between the MBSs at the vortices can be
safely ignored.  However, if $\kappa\sim1$, the hybridization between
MBSs within the vortices can change our results significantly.  From
the basic relations between $B_{\mathrm{c}1}$ and the thermodynamical
critical field $B_{c}$, in the large-$\kappa$-limit, and between $B_c$
and the condensation energy~\cite{tin96}, the above considerations impose a
limiting condition on the dimensionless magnetic-field-dependent
parameter $\beta$ used in our calculations,
\begin{equation}
\beta  
\lesssim \frac{\mathrm{ln}\kappa}{\kappa} \,
\sqrt{\frac{\mathcal{N} [(e\mathrm{V\, nm}^3)^{-1}]}{B_{\mathrm{c}1}
      [\mathrm{T}]}}\quad .
\label{limit_beta}
\end{equation}
Here $\mathcal{N}$ is the normal-state density of states at the Fermi
energy for the superconducting material.
In available materials systems, the right-hand side of
Eq.~(\ref{limit_beta}) can be larger than 1.
Thus our theoretical description is valid for the range of values of
$\beta$ 
shown in Figs.~\ref{fig:didvvsphi} 
and \ref{fig:didvvsphi2}.

In conclusion, we have studied quantum transport in a
superconductor--topological-insulator hybrid structure in the presence
of a perpendicular magnetic field. We have shown that Landau quantization
results in the emergence of a tunable chiral Majorana mode at the edge
of the superconducting region induced on the surface of the
topological insulator. We find that the velocity of this mode can be tuned
by changing the magnitude of the external magnetic field and/or the
chemical potential of the superconductor. The velocity tunability
gives rise to unique signatures in the differential conductance of the
system when the Majorana edge mode is coupled to a normal electronic
lead. Experimental verification of the tunability of the velocity and
the detailed structure of the differential conductance will provide a
new platform to explore Majorana physics.

\section*{Acknowledgment}
We acknowledge financial support by the Swiss SNF, the NCCR Nanoscience, and the
NCCR Quantum Science and Technology.

\section*{References}

\providecommand{\newblock}{}

\end{document}